# A critical review of pre-post surveys designed to measure student epistemology in undergraduate science courses

Kyriaki Chatzikyriakidou

Department of Food Science and Technology, Aristotle University of Thessaloniki, 11km N.R. Thessaloniki - Polugurou, Thermi 57001, Greece; kchatz@afs.edu.gr  (co-first author)

Kristi L. Hall

Office of Undergraduate Education, College of Behavioral & Social Sciences, University of Maryland, College Park, MD 20742, USA; khall@umd.edu (co-first author)

Edward F. Redish

Department of Physics, University of Maryland, College Park, MD 20742, USA (deceased)

Todd J. Cooke

Department of Cell Biology and Molecular Genetics, University of Maryland, College Park, MD 20742, USA; tjcooke@umd.edu (author for correspondence)

## Abstract

The epistemology of science students, i.e., their beliefs about the nature of the knowledge they are learning, about what they have to do to learn it, and about how they will use that knowledge, often plays a powerful role in what they learn in their science courses.

This perspective paper provides a broad overview of the theoretical frameworks, designs, and applications of online pre-post surveys that are available to assess the potential shifts in epistemic perspectives in undergraduate science courses.

We pay particular attention to a recent survey for biology courses called the Maryland Biological Expectation Survey (MBEX).  The MBEX was developed to probe four epistemic themes that are closely aligned with the Vision and Change initiative for reforming undergraduate biology education.

This review is intended to inform STEM teachers about the availability of online epistemological surveys for evaluating the epistemic effects of their courses. These surveys can also help STEM education researchers readily evaluate how different pedagogies, classroom contexts, and other features of learning environments affect the epistemic perspectives of science students.

## Abbreviations
5CCs:   five core concepts

CLASS: Colorado Learning Attitudes Science Survey

DBER:   discipline-based education research

MBEX: Maryland Biology Expectations Survey

MPEX: Maryland Physics Expectations Survey

STEM:   science, technology, engineering, and mathematics

Vision and Change:     V & C

# Introduction – the role of the epistemology of science students in how they learn in their courses

It is well-established in the literature devoted to discipline-based education research (DBER) that the prior experiences of science students affect how they hear and interpret the content we teach in our courses (Adams et al., 2006; Madsen et al., 2015; Redish et al. (1998); Semsar et al., 2011).

Previous schooling will unavoidably influence what is referred to as the epistemic stances of students, i.e., what they expect to happen in our classrooms and how they learn new scientific knowledge.

Epistemology is the branch of classical philosophy that is devoted to the study of knowledge and the related process of knowing that focuses on their natures, beliefs, scopes, validities, and applications.

In the DBER literature, epistemology refers more specifically to our students' beliefs about
- the nature of the knowledge they are learning
- their attitudes about what they have to do to learn it, and
- their expectations about how they will use that knowledge

In essence, the epistemic stances of our students will profoundly shape their willingness and ability to learn disciplinary knowledge and apply it to new learning situations and real-world experiences (e.g., Chatzikyriakidou and McCartney, 2022; Elby, 1999; Gouvea et al., 2019; Hall et al., 2011; Lising & Elby, 2005; Sandoval, 2005; Watkins et al., 2012).

This literature argues that authentic learning experiences can only occur in science classrooms when our students are immersed in the epistemic constructs, values, and beliefs intrinsic to the particular discipline while they are simultaneously learning content knowledge and experimental procedures.

The overall goal of this perspective is to evaluate online survey instruments available to measure pre-post shifts in the epistemic views of undergraduate students taking science courses. The paper has these sections:

- We discuss three theoretical frameworks that can be used to study science student epistemology.
- Then we describe the major pre-post surveys that were designed following those frameworks and review the noteworthy results that have been obtained from those surveys.
- Lastly, we describe the recent Vision and Change (V & C) initiative that has resulted in significant reforms in undergraduate biology education, and we review a new survey instrument that is designed to measure potential shifts in epistemic perspectives occurring in biology students taking V & C-inspired courses.

# Theoretical frameworks for studying science student epistemology

The practitioners of discipline-based education research (DBER) are encouraged to view our research from scientific perspectives that are informed by each discipline's goals, knowledge, and practices (Singer et al., 2012). It follows that we often use explicit theoretical frameworks to help guide that research (Luft et al., 2022; Redish, 2004, 2014). In turn, the resulting observations provide new insights for evaluating and revising the original frameworks. This interplay between theory and experiment contributes to making discipline-based education research (DBER) such a powerful strategy for reforming undergraduate science education (Singer et al., 2012).

Three general frameworks can be applied toward studying science student epistemology:
- The blank-slate framework
- The stages framework
- The resources framework

## Blank-slate framework

The simplest theoretical framework for considering the epistemology of science learning is to view students as being "blank slates," or naive recipients of new knowledge (Locke, 1690). This framework expects that the process by which students acquire new knowledge depends largely on their ability to memorize copious amounts of facts, techniques, and formulas. It assumes a simple and direct type of knowledge transfer from instructors to students that does not incorporate the influences of prior academic, personal, and sociocultural experiences. Because we cannot identify any modern DBER researcher who uses the blank-slate framework as the starting point for studying science student epistemology, we shall not consider this framework any further.

## Stages framework

The stages framework of epistemological development grows out of the cognitive development framework that Swiss psychologist Jean Piaget developed for children (e.g., Piaget and Inhelder, 1958). This framework recognizes the importance of general scientific thinking skills and acknowledges their dependence on students' epistemic stances.

The stages framework of epistemological development for young adults (i.e., college students) predicts that the learning process involves the acquisition of stable and coherent stages of epistemological beliefs and associated disciplinary knowledge that can describe the transition from novice to expert (Perry, 1968; Kitchener and King, 1981; Belenky et al., 1986; Magolda, 1992; Hofer, 2000).

In this framework, epistemological novices may rely on a limited set of epistemological tools and tend to adopt what might be called polar perspectives, meaning that knowledge is something that is either true or not, or something that you either know or don't know. By contrast, experts have access to a broader set of knowledge-building tools, and thus, they see knowledge as being constructed meaning that it is malleable, depends on the context, and may be modified as new information is acquired.

Researchers working in the stages framework tend to assume that attaining a stage is a developmental characteristic of a student that cuts across multiple disciplines (Perry, 1968; Kitchener and King, 1990). Instructors working in the stages framework tend to pay attention to student skill development as well as content knowledge. They might adopt a pedagogy in line with a hierarchical model of thinking skills such as Bloom's taxonomy (Bloom et al., 1956). They might use a scaffolded pedagogy that begins with the accumulation of facts and proceeds to more sophisticated epistemic tools added in coordination with more sophisticated knowledge.

## Resources framework

The third framework we consider here is more finely grained than the stage framework. As an extension of the Knowledge-in-Pieces (KiP) approach developed to analyze student thinking in physics (diSessa, 1993) and math (Sherin, 2001), the resources framework gives us specific ways to analyze student epistemology.

In the resources framework, an individual's knowledge consists of many separate bits (epistemological resources) that the individual considers to be irreducible. ("That's just the way things are.") These resources are learned from experience, including prior schooling. They can become clustered so that they tend to activate together as people become more expert in using them (diSessa and Sherin, 1998; Sabella and Redish, 2007). This framework views students as having these resources that they recruit and use depending on their perception of the context (epistemological frame). In particular, students are not expected to apply the same epistemic resources to every situation.

In more detail, epistemological resources are what we all use to decide that we know something (Hammer, 1994; Hammer and Elby, 2002, 2003). Examples include

- It comes from trusted authority ("It's given in the periodic table.")
- A correct mathematical transformation discloses a true value ("We ran the Punnett square correctly so this should be the correct distribution of genotypes in the F2 generation.")

Epistemological framing (Hammer et al., 2005) refers to the fact that when we come to a new situation, we all make decisions (often not conscious ones) as to what behaviors are appropriate based on our previous physical and social experiences (Goffmann, 1986; Tannen, 1993; Redish, 2014). When coming to a learning experience, students decide what knowledge they are learning and what tools they expect are appropriate to use. Inappropriate choices can create barriers to understanding and resistance to unfamiliar pedagogies.

Researchers working in the resources framework consider not only students' errors and misconceptions, but also the resources from which they could have built the correct responses and appropriate conceptions. These researchers might study the student's perception of a class (epistemic expectations) in a variety of different circumstances and not expect a consistent response. They would look for a shift in probabilities of students' expectations in response to different pedagogical interventions.

Instructors working in a resources framework might want to not adopt a pedagogy that adds epistemic tools in subsequent stages, since starting with simple tools might confirm students' epistemological misconceptions (for example, that memorizing is the main point). These instructors might rather adopt a pedagogy that is both explicit about epistemological issues and mixes epistemic tools from the beginning, showing the value and scaffolding the use of higher-order Bloom stages in simple situations, increasing the complexity of the situation, and decreasing the scaffolding as the class progresses.

In summary, the resources framework differs from the stage framework in that it posits from the start that the epistemologies of even novice students are fragmentary, dynamic, and context dependent. Students can progress from novice to expert, but novices are seen as already possessing some of the resources needed to build expert tools, and experts are seen as sometimes using novice reasoning (and getting incorrect results). Since coherence is an expert epistemological resource, students' epistemic responses in a given course may be variable depending on their epistemological framing.

# Survey instruments available for evaluating science student epistemology

Ultimately, we want to develop more effective methods for characterizing which pedagogical approaches and classroom activities encourage effective student learning in science courses. One aspect of this research was to understand how the personal epistemologies, or epistemic expectations, of science students shape their abilities to learn, integrate, and apply new knowledge gained from their science courses.

An expectation is generally defined as being a metastable and dynamic belief that is potentially sensitive to classroom and laboratory experiences, whereas the term attitude is often restricted to a more stable and deep-seated belief that is relatively unresponsive to the experiences from a single course. These terms are sometimes used interchangeably in the DBER literature (cf. Redish et al., 1998 and Adams et al., 2006).

It is possible to measure students' pre-existing expectations and how a given course might change those expectations via several different methods, including online pre-course (pre) vs. post-course (post) surveys composed of Likert-style statements. Pre-post surveys represent an easy-to-implement approach toward gaining broad-brush impressions about what students think about their attitudes about the knowledge in a particular discipline. They can provide useful insights for teachers about what might be happening in their classrooms, and for researchers about what might be happening across multiple classrooms having different pedagogies, student characteristics, classroom contexts, and other research issues.

In essence, these surveys can serve to identify potential patterns emerging out of the noise of student learning that might be worthy of additional investigation with more focused and rigorous research tools. Thus, surveys can be seen to provide interesting leads, but not definitive interpretations.

Most of the quantitative research on science student epistemology has used two series of epistemological surveys that were designed by two groups of physics education researchers:
- Maryland Expectations Surveys (MEX)
- Colorado Learning Attitudes about Science Surveys (CLASS).

These surveys composed of Likert-style questions are based on different theoretical frameworks, and they attempt to assess similar but not identical aspects of student epistemology. Table 1 shows how they are similar and how they differ.

**Table 1. Key characteristics of selected epistemological surveys**

All information was taken from the latest published versions.

Common statements refer to the number of substantially identical statements found in both the physics and biology surveys from the same group.

| Survey | MPEX | CLASS-Phys | CLASS-Bio | MBEX |
|---|---|---|---|---|
| Theoretical framework | Resources model | Stages model | Stages model | Resources model |
| Design | Pre-post survey with 34 Likert-style statements | Pre-post survey with 42 Likert-style statements | Pre-post survey with 32 Likert-style statements | Pre-post survey with 29 Likert-style statements and 3 open questions |
| Epistemic themes | Independence<br>Coherence<br>Concepts | Real-world connections<br>Personal interest<br>Sense-making/effort<br>Conceptual connections<br>Applied conceptual understanding<br>Problem-solving: General<br>Problem-solving: Confidence<br>Problem-solving: Sophistication | Real world connections<br>Enjoyment (personal interest)<br>Problem-solving: Reasoning<br>Problem-solving: Synthesis and application<br>Problem-solving: Strategies<br>Problem-solving: Effort<br>Conceptual connections /memorization | Principles<br>Independence<br>Interdisciplinary reasoning<br>Relevance |
| Common statements | 6 | 19 | 19 | 6 |
| References | Redish et al. (1998)<br>McCaskey (2009) | Adams et al. (2006)<br>PhysPort (2025) | Semsar et al. (2011)<br>SEI (n.d.) | Hall (2013)<br>Hall et al. (2024) |

In this section, we first review the physics versions of these surveys because the extensive work on physics student epistemology has frequently used these surveys. We then discuss the biology versions because we believe that they will need more revision to be broadly applicable to life-sciences courses that are designed according to the guidelines in the *Vision and Change*

reports (AAAS, 2011, 2015, 2018).

Other surveys for measuring student expectations in physics (Halloun and Hestenes, 1998; Elby, 2001) and chemistry (Barbera et al., 2008; Grove and Bretz, 2007) will not be discussed here.

## Maryland Physics Expectations Survey (MPEX)

Redish and his colleagues (1998) developed, validated, and used the Maryland Physics Expectations Survey (MPEX) to study student expectations about the process of learning physics. Both MPEX versions used the resources framework of student epistemology, which proposes that disciplinary knowledge of students and their attitudes about that knowledge are composed of fragmentary, often inconsistent, and context-dependent assemblages.

Epistemic clusters (= categories or themes) in the MPEX series were identified a priori by disciplinary experts as being important clusters that contribute to how students learn in introductory science courses. Individual statements in the different clusters were said to be "non-orthogonal," meaning that they did not necessarily fit exclusively into a single cluster but rather often overlapped across several clusters.

Their interviews and validation studies convinced the developers that the statements on an expectations survey needed to be specific to the population of students, as many of the students' expectations grew out of their previous classroom experiences and their perception of the nature of their discipline.

The original version of the MPEX (MPEX-1) was developed to enable broad surveys of student epistemological framings within the context of the calculus-based introductory physics sequence for mostly engineering students. Redish et al. (1998) developed the survey based on extensive observations of student performance and numerous follow-up interviews.

The MPEX has 34 Likert-style statements exploring 6 overlapping themes (called clusters): independence, coherence, concepts, reality link, math link, and effort. They delivered the survey to ~1,500 students at 3 large public universities.

To expand their studies to a larger population, the original MPEX was modified to create MPEX-2 specifically designed for students taking introductory algebra-based physics courses that included many biology and other life-science majors (McCaskey, 2009). This effort was a component of a project to study and develop instructional materials that fostered student epistemological development (Redish and Hammer, 2009). MPEX-2 clarified certain statements for these students resulting in fewer misinterpretations, eliminated the effort cluster (which has ambiguous implications), and grouped the reality-link and math-link statements into the coherence and concepts clusters, respectively.

Because meta-analyses often lump together the results from the MPEX and the CLASS-Physics

Student epistemology in science coursessurveys, our discussion of the MPEX results is deferred until after we introduce the CLASS-Physics survey.

## Colorado Learning Attitudes about Science Survey - Physics (CLASS-Phys)

The Colorado Physics Education Research Group developed the CLASS series of epistemological surveys, including CLASS-Physics (Adams et al., 2006), CLASS-Chemistry (Barbera et al., 2008), and CLASS-Biology (Semsar et al., 2011).

The CLASS surveys also used Likert-like statements that were grouped into overlapping categories similar to MEX clusters. The CLASS surveys were explicitly constructed from the perspective of distinguishing the expert attitudes of disciplinary practitioners from the naive expectations of students. This approach of identifying a novice-to-expert spectrum implicitly assumed the stage framework of student epistemology. which proposes that the learning process involves the acquisition of stable and coherent stages of disciplinary knowledge and associated epistemological attitudes that follow the transition from novice to expert.

CLASS-Phys (Adams et al., 2006) differs from the MPEX in several other fundamental ways. Along with incorporating important epistemic categories from expert opinion, other categories were empirically determined from student responses by using principal component analysis to identify those exhibiting some coherence. This approach toward survey construction offers the advantage that student attitudes underlying these categories may be more readily addressable in the classroom. However, due to the likely incoherent nature of naïve student epistemology, this might also eliminate some categories worthy of further investigation.

Two, student responses were classified into 8 overlapping categories: real-world connections, personal interest, sense-making/effort, conceptual connections, applied conceptual understanding, problem-solving general, problem-solving confidence, and problem-solving sophistication. Thus, the range of measured categories extended beyond the boundaries of what would ordinarily be considered as epistemological to include affective issues and disciplinary practices. Lastly, as was the case with the MEX series, an individual statement in the CLASS series could also be used to assess to student thinking about several different themes.

The original MPEX paper reported that student expectations tended to deteriorate over the first semester of calculus-based introductory physics (Redish et al., 1998). Follow-up studies using the MPEX and CLASS-Phys at many other universities confirmed this result in traditional large lecture classes, even ones that produced measurable gain in conceptual understanding (Adams et al., 2006; Madsen et al., 2015).

However, certain pedagogical interventions and classroom contexts were able to induce reproducible positive gains in student expectations in physics courses at various universities (McCaskey 2009; Redish and Hammer, 2009; Lindsey et al., 2015; Madsen et al., 2015).

Page 9 of 23

Typically, the classes that showed significant gains were designed to focus on epistemological issues. In particular, the method that organizes around modeling (Hestenes, 1987) was effective in large introductory physics classes (Brewe et al., 2009, 2013), and workshop-style active-engagement instruction was very effective in obtaining improved survey results in both large introductory courses (Gatch, 2010) and small classes for service teachers (Otero and Gray, 2009).

### Colorado Learning Attitudes about Science Survey - Biology (CLASS-Bio)

CLASS-Bio was designed to probe 7 overlapping categories, similar to but not identical to those identified in the CLASS-Phys: real world connections, enjoyment (personal interest), problem-solving: reasoning, problem-solving: synthesis and application, problem-solving: strategies, and problem-solving: effort (Semsar et al., 2011). Nineteen of the 32 Likert statements in CLASS-Bio were taken directly or modified slightly from CLASS-Phys and CLASS-Chem. The remaining 13 unique statements were generated by faculty working groups, student interviews, faculty reviews, and factor analysis (Semsar et al., 2011: Table 1).

A genuine advantage of CLASS-Bio is that the multi-departmental, multi-institutional process used to design CLASS-Bio makes it valid for assessing biology courses offered at all levels from introductory to upper-division courses across the full range of undergraduate institutions.

Because the entire series of CLASS surveys shared a common theoretical framework and design process that resulted in a majority of equivalent Likert statements in all 3 surveys, the CLASS-Bio is relatively neutral with respect to scientific discipline. Therefore, it offers the additional benefit of potentially being useful for surveying biology students taking required introductory physics and chemistry courses.

In its first application, the CLASS-Bio showed that introductory biology courses often resulted in significant shifts toward less expert-like student attitudes (Semsar et al., 2011), just like the initial results from other expectation and attitude surveys (Redish et al., 1998; Adams et al., 2006; Barbera et al., 2008). Using the CLASS-Bio, Ding and Mollohan (2015) observed similar declines in the epistemic perspectives of science majors taking an introductory biology course but slight increases in the epistemologies of non-science students taking the non-majors version. Students enrolled in typical upper-division courses did not exhibit similar declines but would sometimes show significant positive gains (Hansen and Birol, 2014; Elliott et al., 2016; Beumer, 2019).

Meaningful improvements in biology student attitudes have been observed in individual courses as the result of various pedagogical innovations, such as active-learning approaches (Cleveland et al., 2017), peer tutoring (Batz et al., 2015), scale-up classrooms (Soneral and Wyse, 2017), course-based undergraduate research experiences (Olimpo et al., 2016; Pavlova et al., 2021), and the CREATE (Consider, Read, Elucidate hypotheses, Analyze and interpret the data, Think of the next Experiment) strategy (Hoskins and Gottesman, 2018).

In a seminal study, Hansen & Birol (2014) used CLASS-Bio to characterize student attitudes during the entire four years of an undergraduate biology program at a major research university. The group of matched respondents who took the CLASS-Bio in both their first and fourth years exhibited a positive overall gain from 64.5 to 72% more expert-like responses, with the largest improvements in the real-world connection and enjoyment (personal interest) categories.

However, it is worth noting that because CLASS-Bio was largely derived from a pre-existing physics epistemology survey, it was not designed to assess the epistemic stances of biology students enrolled in new reformed courses that follow the guidelines proposed in the *Vision and Change* initiative.

## *Vision and Change* (V & C) initiative for reforming undergraduate biology education

Since 2000, multiple reports have argued for transforming the education offered to undergraduate biology and pre-medical students (e.g., Bio2010 (NRC, 2003), *Scientific Foundations for Future Physicians* (AAMC-HHMI, 2009), *A New Biology for the 21st Century* (NRC, 2009).

This broad effort to transform undergraduate biology education culminated in the *Vision and Change* (V & C) initiative (AAAS, 2011, 2015, 2018). This initiative laid out an explicit blueprint for transforming undergraduate biology education, which focused on the importance of

- integrating five core concepts (5CCs) and scientific competencies throughout the undergraduate biology curriculum
- changing biology pedagogy to focus on student-centered learning
- promoting campus-wide commitments to meaningful change, and
- engaging the entire biology community in implementing that change

The 5CCs were identified as:
- Evolution
- Information Flow, Exchange, and Storage
- Structure and Function
- Pathways of Transformation of Energy and Matter, and
- Systems.

The conceptual framework constructed from the 5CCs and scientific competencies promises to offer a coherent and integrated approach for helping undergraduate biology students learn, organize, and apply the fundamental knowledge and experimental skills needed to become practicing biologists (e.g., Branchaw et al., 2020; Chatzikyriakidou et al., 2021; Couch et al., 2019).

The V & C initiative has led to major curricular improvements in the undergraduate biology education. Couch et al. (2019) developed an assessment instrument called the General Biology–Measuring Achievement and Progression in Science (GenBio-MAPS) that measures how well biology students understand the 5CCs from V & C at three key time points in their undergraduate biology programs. These authors found that biology students at 20 institutions showed significant trends in improved understanding of most, but not all, ideas from the 5CCs over the course of their undergraduate studies. Introductory biology students working in small groups on active-learning activities were shown to develop a deep appreciation of the V & C competency that biology should be practiced as an interdisciplinary science (Jardine et al., 2017). Chatzikyriakidou and McCartney (2020) reported that student understanding of the 5CCs was positively correlated with more favorable epistemology scores measured at the end of the course.

Moreover, it is widely accepted that pedagogical strategies promoting active learning (also called active engagement have the potential to encourage self-directed learning and critical thinking in undergraduate STEM students (for overviews, see Michael, 2006; Wood, 2009; Singer *et al.*, 2012; Mintzes and Walter, 2020; Yannier *et al.*, 2021.) In a meta-analysis across STEM disciplines, including biology, Freeman *et al.* (2014) documented that student performance on exams was significantly higher in courses using student-centered strategies as opposed to traditional lectures. One caveat from this analysis is that the positive impacts were most evident in small classes of $\leq$ 50 students, whereas introductory biology courses often have higher enrollments, especially at large public and private universities.

Lastly, student-centered pedagogies can significantly increase equity and inclusion in STEM education: High-intensity active engagement caused a large reduction in the achievement gaps for underrepresented students who might feel less supported by traditional lecturing (Theobald et al., 2020).

A later Science editorial urged the renewed commitment to the V & C initiative (Asai et al., 2022).

Because the often-observed correlation between content learning and student epistemologies described in the Introduction cannot be taken to imply a causal relationship, both curricular material and assessment tools have to be designed in ways that reflect the disciplinary epistemic knowledge that majors need to acquire during their degree programs.

Of particular relevance to the development of a biology epistemological survey is the issue of what epistemic themes should be emphasized to align with the V & C initiative. From the onset, it must be acknowledged that the word epistemology or any of its variants is never explicitly mentioned in the original V & C report (AAAS, 2011). Yet, this reform initiative is deeply epistemological in that it focuses on the nature of biological knowledge and the process of learning that knowledge.

Hall (2013) identified four major epistemic themes that are implicitly stressed in the V & C report:

- **Principles**: Expectations about the nature of biological knowledge
- **Independence**: Expectations about the process of learning biology
- **Interdisciplinary reasoning**: Expectations about the value of incorporating other scientific disciplines into biology education
- **Relevance**: Expectations about the purpose of biology education

## Maryland Biology Expectations Survey (MBEX)

The effort to reform undergraduate biology education that culminated in the *Vision and Change* initiative (AAAS, 2011, 2015, and 2018) motivated a large US public university to transform its entire introductory biology curriculum. Particular attention was devoted to the organismal biology course in the introductory biology sequence that covers the diversity and functions of all organisms (Cooke and Jensen, 2022; Cooke et al., 2023).This course was specifically revised to emphasize fundamental principles, multidisciplinary perspectives, and quantitative reasoning.

The MBEX was created to evaluate the effects of different pedagogical approaches and classroom contexts on the epistemology of biology students taking that course (Hall, 2013; Hall et al., 2024). Just like MPEX, this pre-post survey was based on the resources framework because semi-structured interviews with biology students, both in introductory organismal biology courses and introductory physics classes designed for life-science students, revealed that their epistemic perspectives were often fragmentary, context-dependent, and internally inconsistent (Hall et al., 2011; Watkins et al., 2012; Watkins and Elby, 2013; Sawtelle and Turpen, 2016; Gouvea et al., 2019).

A striking example of the dynamic nature of student epistemology came from a semi-structured interview exploring a student's ideas about using physics equations in this introductory organismal biology course (Hall et al., 2011; Watkins and Elby, 2013). In the first part of the interview, she expressed what appeared to be a fixed and hostile attitude toward physics equations, framing math in biology as irrelevant and unpleasant. She felt that she had to memorize equations for exams but saw them as having no value in sense making.

Yet, later in the same interview, she became excited when she recounted classroom demonstrations on the relationship between surface area to volume in biological scaling. Then, she framed the use of math as having authentic value to biological thinking. She found it "mind-boggling" that the mathematics of scaling dictated that for model wooden horses showing two stages of isometric growth, the larger horse would inevitably and dramatically collapse under its own weight.

Using the resources framework, the MBEX was designed and validated for measuring four epistemic themes of principles, independence, interdisciplinary reasoning, and relevance that are related to the higher-level skills emphasized in the forementioned reports on reforming biology education (Hall, 2013; Hall et al., 2024). For the students taking the introductory organismal course described above, the MBEX revealed that active-engagement pedagogy, but not conventional lectures, tended to significantly improve their expectations for the themes of principles and interdisciplinary reasoning (Hall, 2013). Post-course responses revealed that specific pedagogies used in different versions did not dislodge their entrenched attitudes toward their preferred pedagogical style. Regardless of the pedagogy used, all versions of this course resulted in significant deterioration of favorable pre-class attitudes toward the relevance theme. The MBEX was thus able to provide useful insights for future efforts for improving the epistemic stances of introductory organismal biology students.

Chatzikyriakidou and McCartney (2020) distributed pre- and post- MBEX surveys in a large general biology course offering 3 lecture, 1 discussion, and 1 laboratory sessions each week. These authors observed that students expressed slight, albeit insignificant, declines in all 4 epistemic themes, which is similar to the discouraging results obtained from epistemological surveys offered in other lecture-based introductory courses in both physics and biology (Adams et al., 2006; Ding and Mollahan, 2015; Madsen et al., 2015; Redish et al., 1998; Semsar et al., 2011). Using a more fine-grained analysis focusing on the responses of students to individual statements, Chatzikyriakidou and McCartney (2020) were able to identify potential mismatches in students' attitudes about how they view science as a discipline and how they learn biology in a conventional introductory biology course.

In particular, these students exhibited favorable responses toward general science statements, while they were simultaneously expressing unfavorable attitudes toward more specific statements about learning biology. The observed mismatch in this course between their general attitudes about science and their specific expectations about learning biology is probably linked to their impressions of the learning environment in a conventional biology course.

The MBEX survey seems like a worthwhile starting place for developing an effective biology epistemology survey that can be broadly applied to reformed undergraduate biology courses.

One notable strength of the MBEX is that its design is based on the resources framework, which seems to more accurately describe how biology students express their attitudes toward biology knowledge and the process of learning it. Moreover, since the MBEX-2 was specifically developed to align with the epistemic themes presented in the V & C initiative, it can be used to assess whether V & C-inspired reforms in course content and pedagogical strategy are having favorable effects on student epistemologies.

However, more effort research is needed on the design and validation of the MBEX before it can be reliably used to survey other biology courses at other universities. The interpretation of the survey statements of assigned to each epistemic theme in MBEX-2 should be reevaluated by using structured interviews with both faculty and students in classroom contexts other than the organismal biology course used to develop this survey.

Indeed, the initial factor analyses of Chatzikyriakidou and McCartney (2022) suggested that the MBEX might actually address the following four epistemic themes:

- interdisciplinary knowledge
- application of interdisciplinary knowledge
- real-world connections = relevance, and
- learning facts v. principles in class.

This structure overlaps with the original one presented in Hall (2013), except that the epistemic theme of independence regarding the process of learning biology did not emerge from their factor analyses.

It is worth pointing out that factor analysis tends to identify those aspects of student thinking having some coherence, thereby reducing the variability in student responses (Semsar et al., 2011), whereas the resources framework does not necessarily expect student responses to exhibit such coherence. One could instead argue that epistemic incoherence might reveal interesting attitudes worth further study. Nevertheless, since the independence theme is emphasized in V & C report, this work suggests that the statements assigned to that theme in the current MBEX should be carefully evaluated before constructing a revised MBEX.

## Conclusions

The epistemology of science students, i.e., their attitudes toward scientific knowledge and the process of learning it, has profound effects on their willingness and ability to learn in their science courses.

Two groups working in DBER have used different theoretical frameworks to develop similar online pre-post surveys to measure science student epistemology.

This review summarizes the effort to develop a new biology epistemology survey called the MBEX that is closely aligned to the V & C initiative in undergraduate biology education.

Once the MBEX is further validated for different biology courses across multiple institutions, it is expected that this survey should become an effective tool for both classroom instructors and education researchers for assessing how well their biology courses are addressing and teaching the epistemic themes of the V & C initiative.

## Acknowledgments

We are grateful for the support, guidance, and collegiality from the Physics and Biology Education Research Groups at the University of Maryland who have helped us to design, validate, and implement the MPEX and MBEX surveys for more than 25 years. All work on developing the MBEX survey was approved by the University of Maryland Institutional Review Board (IRB #09-0080). Our MBEX research was supported in part by NSF grant DUE 0919816 to TJC and EFR.

Student epistemology in science courses

Student epistemology in science courses
nap.nationalacademies.org/catalog/13362/discipline-based-education-research-understanding-and-improving-learning-in-undergraduate (accessed 1 July 2025).

Soneral, P. A. G. & Wyse, S. A. (2017). A SCALE-UP mock-up: comparison of student learning gains in high- and low-tech active-learning environments. *CBE-Life Sciences Education,* 16(2), ar12. doi:10.1187/cbe.16-07-0228

Tannen, D. (1993) *Framing in discourse*. (New York, Oxford University Press).

Theobald, E. J., Hill, M. J., Tran, E., Agrawal, S., Arroyo, E. N., Behling, S., ... & Freeman, S. (2020) Active learning narrows achievement gaps for underrepresented students in undergraduate science, technology, engineering, and math. *Proceedings of the National Academy of Sciences (USA)*, 117(12), 6476-6483. doi:10.1073/pnas.1916903117

Watkins, J. & Elby, A. (2013) Context dependence of students' views about the role of equations in understanding biology. *CBE-Life Sciences Education*, 12(2), 274-286. doi:10.1187/cbe.12-11-0185

Watkins, J., Coffey, J. E., Redish, E. F., & Cooke, T. J. (2012). Disciplinary authenticity: Enriching the reforms of introductory physics courses for life-science students. *Physical Review Special Topics—Physics Education Research, 8*(1), 010112. doi:10.1103/PhysRevSTPER.8.010112

Wood, W. B. (2009) Innovations in teaching undergraduate biology and why we need them. *Annual Review of Cell and Developmental Biology*, 25, 93-112. doi:10.1146/annurev.cellbio.24.110707.175306

Yannier, N., Hudson, S. E., Koedinger, K. R., Hirsh-Pasek, K., Golinkoff, R. M., Munakata, Y., ... & Brownell, S. E. (2021) Active learning: "hands-on" meets "minds-on". *Science* 374(6563), 26-30. doi:10.1126/science.abj9957

Page 23 of 23